\documentstyle[preprint,prb,aps,epsf]{revtex}
\begin{document}
\draft
\title{ Interlayer coupling and
the $c$-axis quasiparticle transport in high-$T_{c}$ cuprates}
\author{Wonkee Kim and J. P. Carbotte}
\address{Department of Physics and Astronomy, McMaster University,
Hamilton, Ontario, Canada L8S~4M1}
\maketitle
\begin{abstract}
The $c$-axis quasiparticle conductivity shows different behavior
depending on the nature of the interlayer coupling. 
For coherent coupling with a constant hopping
amplitude $t_{\perp}$, the conductivity at zero frequency and
zero temperature $\sigma(0,0)$
depends on the direction of the magnetic field, but
it does not for angle-dependent hopping $t(\phi)$ which
removes the contribution of the nodal quasiparticles.
For incoherent
coupling, the conductivity is also independent of field direction
and changes only when paramagnetic effects are included.
The conductivity sum rule can be used to determine the admixture
of coherent to incoherent coupling. 
The value of $\sigma(0,0)$ can be dominated by $t_{\perp}$ while
at the same time $t(\phi)$ dominates the
temperature dependence of the superfluid density.
\end{abstract}

\section{introduction}
The nature of the interlayer coupling between two adjacent CuO$_{2}$ planes
in the cuprates is an important issue that remains unresolved.
Suggestions include effects of strong intralayer scattering,
\cite{kumar} non-Fermi liquid ground states,\cite{anderson1}
the general phenomenon of confinement,\cite{anderson2}
inter- and in-plane charge fluctuations,\cite{legget}
indirect $c$-axis coupling through the particle-particle channel,
\cite{tesanovic} as well as resonant tunneling on localized states
in the blocking layer\cite{halbritter} and two band models.\cite{atkinson}
Coherent coupling originates from an overlap of the electronic
wave functions between planes, and in-plane momentum is conserved
in interlayer hopping. By contrast for
impurity-mediated incoherent coupling, the in-plane momentum
is not constraint.
It has been shown that the $c$-axis conductivity sum
rule depends on the nature of interlayer coupling.\cite{kim1}
Coherent coupling obeys the conventional sum rule\cite{fgt} regardless of 
the angular dependence of the interlayer hopping amplitude.
On the other hand, incoherent coupling violates the sum rule
even if the in-plane dynamics can be described by a Fermi liquid.
However, in order to 
explain the violation observed in some experiments,\cite{basov} 
it was also necessary to include
the non-Fermi liquid nature of the in-plane dynamics.\cite{kim2}
For YBa$_{2}$Cu$_{3}$O$_{7-\delta}$ at optimum doping, a conventional
$c$-axis sum rule is observed\cite{basov} which is consistent
with coherent $c$-axis coupling and an in-plane Fermi liquid.
For the underdoped case, a pseudogap is observed and the sum rule
is closer to $1/2$.\cite{basov,ioffe} 
This value can most easily be understood
as a pseudogap effect with incoherent 
$c$-axis coupling.\cite{kim2}

Another interesting example of the interlayer coupling
is the $c$-axis conductivity due to
low lying quasiparticles at zero temperature and zero
frequency. A method for observing the quasiparticle current
is to measure the hysteretic $c$-axis tunneling current-voltage
characteristics of layered cuprates.\cite{latyshev}
For measurements of the optical conductivity which includes
the pair tunneling, see Ref.\cite{homes}
For coherent coupling with a constant hopping amplitude $t_{\perp}$,
the $c$-axis quasiparticle conductivity $\sigma(w\rightarrow0,
T\rightarrow0)$ shows a residual value independent of the in-plane
scattering rate $\gamma \ll\Delta_{0}$ (gap) to leading order 
\cite{latyshev} with the correction $-(\gamma/\Delta_{0})^{2}$.
The underlying physics in this case is the same as for the in-plane 
quasiparticle conductivity;\cite{lee,durst,inplane} 
namely, the impurity-induced density
of states at $\omega=0$ is canceled by the decrease of quasiparticle lifetime.
A universal value has been observed in
YBa$_{2}$Cu$_{3}$O$_{6.9}$ for the in-plane thermal conductivity.\cite{univ} 
In some sense the thermal conductivity is an ideal probe of universal
behavior since, as opposed to the electrical conductivity,
it is not renormalized by vertex and by Fermi-liquid corrections,\cite{durst}
and so there is less ambiguity in its identification.
The universal conductivity limit does not appear for
coherent coupling with an angle-dependent hopping amplitude of the form
$t(\phi)=t_{\phi}\cos^{2}(2\phi)$ which is believed
to be appropriate for the copper oxides, 
where $\phi=\tan^{-1}(k_{y}/k_{x})$ is an angle
in the momentum space. In this case the $c$-axis quasiparticle 
conductivity is reduced
by a factor of $(3/8)\left(\gamma/\Delta_{0}\right)^{2}$ as
compared with the value of
$\sigma(w\rightarrow0,T\rightarrow0)$ for a constant $t_{\perp}$
assuming the same magnitude of hopping amplitude, which of course
is not the case.
This arises because the angular dependence, $\cos^{2}(2\phi)$, 
eliminates the
contribution of the quasiparticles on the nodal lines from interlayer
transport. For incoherent coupling, the residual conductivity
is proportional to $\Bigl(\gamma\ln[\Delta_{0}/\gamma]/\Delta_{0}
\Bigr)^{2}$ in leading order. Consequently, the universal
value of the $c$-axis quasiparticle conductivity is 
a characteristic only of coherent coupling with a constant
hopping amplitude. 
Geometrical consideration of the Cu and O atomic arrangements\cite{sand,xiang}
from plane to plane leads one to expect that the dominant overlapping
of orbitals would lead to a $t(\phi)$ form with possibly 
a subdominant constant
piece $t_{\perp}$. In this case, $t_{\perp}$ would still dominate
the value of $\sigma(0,0)$ for very small $\gamma$, but the other terms
would become important as $(\gamma/\Delta_{0})^{2}$ corrections
become larger in which case the term $t(\phi)$ or
the incoherent part can become important.

In this paper we investigate effects of interlayer couplings on
the $c$-axis quasiparticle transport in the absence, 
as well as in the presence of 
an in-plane magnetic field.
It is our aim in this paper to look at all three contributions
of interlayer coupling in detail,
and our discussion of the magnetic field effects will not be restricted
to the constant $t_{\perp}$ as it is in Ref.\cite{bulaevskii}
We also investigate the role played by the conductivity sum rule in
quasiparticle transport.
We find that it can be used to fix the relative amount of
coherent to incoherent coupling.
We also find that $\sigma(\omega,0)/\sigma(0,0)$
shows a different $\omega^{2}$ coefficient for
$\omega<\gamma$ and $\sigma(0,T)/\sigma(0,0)$ a different $T^{2}$
coefficient $(T<\gamma)$
depending on the nature of the coupling. 
In the presence of an in-plane magnetic field,
$\sigma_{\bf q}(0,0)$ for coherent coupling with a constant hopping 
amplitude $t_{\perp}$
depends on the direction of the field\cite{bulaevskii}
as does the coefficient of the $T^{2}$ term in $\sigma_{\bf q}(0,T)$,
but such a dependence is negligible for an
angle-dependent coherent $t(\phi)$ because
it removes the contribution from the nodal quasiparticles which
otherwise manifest the effect of field direction.  
For incoherent
coupling, $\sigma_{\bf q}(0,0)$ is independent of field direction,
and in fact changes only when a paramagnetic interaction is included.
Here we mention that there exist, in the literature, studies which
consider effects of oxigen doping on the $c$-axis transport and
Coulomb charging effect\cite{halbritter} associated with
a pseudogap behavior.\cite{krasnov} Such complication, however,
is beyond the scope of the present paper.

This paper is organized as follows: In Sec. II, we derive general 
formulas associated with the $c$-axis quasiparticle conductivity
including paramagnetic effects. 
We describe, in Sec. III, that effects of coherent coupling
with a constant as well as with an angle-dependent hopping amplitude on
the $c$-axis quasiparticle transport with and without
an in-plane magnetic field.
In Sec. IV, impurity-mediated incoherent $c$-axis
coupling is considered. We also illustrate the
role of the conductivity sum rule in quasiparticle transport
in Sec. V
and draw conclusions in Sec. VI.

\section{formalism}
The Hamiltonian $H$ for a cuprate superconductor with interlayer
coupling can be written as $H=H_{0}+H_{c}$, where $H_{0}$ describes
a $d$-wave superconductor and 
\begin{equation}
H_{c}=\sum_{\sigma,{\bf k},{\bf p}}\left[t_{{\bf k}-{\bf p}}
C^{+}_{1\sigma}({\bf k})C_{2\sigma}({\bf p})+h.c\right]\;.
\end{equation}
The interlayer hopping amplitude $t_{{\bf k}-{\bf p}}$ depends on
nature of $c$-axis couplings:
i) $t_{{\bf k}-{\bf p}}=t_{\perp}\delta_{{\bf k}-{\bf p}}$
for coherent coupling with a constant amplitude. ii) For
an angluar dependent amplitude
$t_{{\bf k}-{\bf p}}=t_{\phi}\delta_{{\bf k}-{\bf p}}\cos^{2}(2\phi)$.
In the lattice, it can be seen from geometrical consideration\cite{sand}
that
$t_{\phi}\delta_{{\bf k}-{\bf p}}
\left[\cos(k_{x}a)-\cos(k_{y}a)\right]^{2}$
with an in-plane lattice constant $a$.
ii) For incoherent coupling $t_{{\bf k}-{\bf p}}=V_{{\bf k}-{\bf p}}$ and
an impurity average is to be taken into account.

Applying perturbation theory for $H_{c}$, we obtain the $c$-axis
quasiparticle conductivity $\sigma(\omega,T)$ in terms of 
the electronic spectral
weight function $A({\bf k},\epsilon)$.
\begin{equation}
\sigma(\omega)=-{C\over\omega}\sum_{{\bf k},{\bf p}}
t^{2}_{{\bf k}-{\bf p}}
\int d\epsilon\left[f(\epsilon+\omega)-f(\epsilon)\right]
A({\bf k},\epsilon)A({\bf p},\epsilon+\omega)\;,
\end{equation}
where $C$ is a constant which depends on 
the nature of interlayer coupling, $f(\epsilon)$ is the usual
Fermi thermal factor, and
the spectral function
\begin{equation}
A({\bf k},\epsilon)={\gamma(1+\xi_{\bf k}/E_{\bf k})\over
{(\epsilon-E_{\bf k})^{2}+\gamma^{2}}}+
{\gamma(1-\xi_{\bf k}/E_{\bf k})\over
{(\epsilon+E_{\bf k})^{2}+\gamma^{2}}}
\end{equation}
with $E_{\bf k}=\sqrt{\xi^{2}_{\bf k}+\Delta^{2}_{\bf k}}$.
We assume 
$\Delta_{\bf k}=
\Delta_{0}\left[\cos(k_{x}a)-\cos(k_{y}a)\right]$
(or $\Delta_{0}\cos(2\phi)$ in the
continuum limit) and a cylindrical Fermi surface with
$\xi_{\bf k}=k^{2}/(2m)-\epsilon_{F}$ measured from
the Fermi energy $\epsilon_{F}$.
As $T\rightarrow0$ and $\omega\rightarrow0$,
$\sigma(0,0)=C\sum_{{\bf k},{\bf p}}t^{2}_{{\bf k}-{\bf p}}
A({\bf k},0)A({\bf p},0)$. It is easy to calculate $\sigma(0,0)(\equiv
\sigma_{0})$ for the couplings we mentioned earlier. 
To obtain the $c$-axis
quasiparticle conductivity at
finite frequency $\omega$ at zero temperature, we take
$T\rightarrow0$ limit and assume $\omega<\gamma$
\begin{eqnarray}
\sigma(\omega)\simeq&&-C\sum_{{\bf k},{\bf p}}t^{2}_{{\bf k}-{\bf p}}
\int d\epsilon\left[\partial_{\epsilon}f(\epsilon)+
\left({\omega\over2}\right)
\partial^{2}_{\epsilon}f(\epsilon)
+\left({\omega^{2}\over6}\right)\partial^{3}_{\epsilon}f(\epsilon)
\right]
A({\bf k},\epsilon)A({\bf p},\epsilon+\omega)
\nonumber\\
\simeq&&C\sum_{{\bf k},{\bf p}}t^{2}_{{\bf k}-{\bf p}}
A({\bf k},0)A({\bf p},\omega)
-C\sum_{{\bf k},{\bf p}}t^{2}_{{\bf k}-{\bf p}}
\left({\omega\over2}\right)\partial_{\epsilon}A({\bf k},\epsilon)
A({\bf p},\epsilon+\omega)\Big|_{\epsilon=0}
\nonumber\\
+&&\; C\sum_{{\bf k},{\bf p}}t^{2}_{{\bf k}-{\bf p}}
\left({\omega^{2}\over6}\right)\partial^{2}_{\epsilon}
A({\bf k},\epsilon)A({\bf p},\epsilon+\omega)\Big|_{\epsilon=0}+\cdots
\label{finiteo}
\end{eqnarray}
where an expansion in $\omega$ has been made.
The finite $T$ corrections at $\omega=0$ will be considered later along
with an in-plane magnetic field.

In the presence of an in-plane magnetic field, we
assume that the field penetrates freely into the sample
so that the field is uniform between the planes.\cite{bulaevskii} 
This assumption should
be valid for Bi- and Tl-based cuprates which are nearly two dimensional. 
For YBCO, however,
there is a possibility that in-plane vortices form. This means that
in such a case an average over vortices in a unit cell is 
required.\cite{vortex}
The interlayer coupling Hamiltonian is modified by the presence of a
uniform in-plane magnetic field ${\bf H}$ as follows:
\begin{equation}
H_{c}=\sum_{\sigma,{\bf k},{\bf p}}\left[t_{{\bf k}-{\bf p}}
C^{+}_{1\sigma}({\bf k}+{\bf q})C_{2\sigma}({\bf p})+h.c\right]\;,
\end{equation}
where ${\bf q}=(ed/2)({\hat {\bf z}}\times{\bf H})$ 
with an interlayer distance $d$.

Following a procedure similar to the one we applied in the zero field case,
we obtain the $c$-axis
quasiparticle conductivity $\sigma_{\bf q}(\omega,T)$ in the presence of
an in-plane magnetic field,
\begin{equation} 
\sigma_{\bf q}(\omega)=-{C\over\omega}\sum_{{\bf k},{\bf p}}
t^{2}_{{\bf k}-{\bf p}}
\int d\epsilon\left[f(\epsilon+\omega)-f(\epsilon)\right]
A({\bf k}+{\bf q},\epsilon)A({\bf p},\epsilon+\omega)\;.
\end{equation}
In our consideration, the energy scale associated with the magnetic field
are always much less than the gap $\Delta_{0}$.
Due to an in-plane field, the quasiparticles gain an additional
momentum ${\bf q}$ on transferring to the next plane. 
In other words, the Bogoliubov-de Gennes
(BdG) wave functions\cite{bdg} 
$u_{\bf k}({\bf r})$ and $v_{\bf k}({\bf r})$
become $u_{\bf k}e^{i({\bf k}+{\bf q})\cdot{\bf r}}$ and
$v_{\bf k}e^{i({\bf k}+{\bf q})\cdot{\bf r}}$, respectively.\cite{fflo}
The spectral weight function becomes
\begin{equation}
A({\bf k}+{\bf q},\epsilon)=
{\gamma\left(1+\xi_{{\bf k}+{\bf q}}/E_{{\bf k}+{\bf q}}\right)\over
\left(\epsilon-E_{{\bf k}+{\bf q}}\right)^{2}+\gamma^{2}}+
{\gamma\left(1-\xi_{{\bf k}+{\bf q}}/E_{{\bf k}+{\bf q}}\right)\over
\left(\epsilon+E_{{\bf k}+{\bf q}}\right)^{2}+\gamma^{2}}\;, 
\end{equation}
where 
$E_{{\bf k}+{\bf q}}\simeq\sqrt{\xi^{2}_{{\bf k}+{\bf q}}+\Delta^{2}_{\bf k}}$
with $\xi_{{\bf k}+{\bf q}}\simeq\xi_{\bf k}+({\bf k}\cdot{\bf q})/m$.
Note that the angular dependence of $\sigma_{\bf q}$
comes from the factor $({\bf k}\cdot{\bf q})/m
\simeq E_{q}\cos(\phi-\theta_{q})$,
where $E_{q}=v_{F}q$ and 
$\theta_{q}$ is the direction of ${\bf q}$, which can be interpreted
as the direction of the field $(\theta)$
because of the symmetry in the problem.
The gap seen by quasiparticles with a momentum ${\bf k}+{\bf q}$ 
is not $\Delta_{\bf k}$ but $\Delta_{{\bf k}+{\bf q}}$; however,
$\Delta_{{\bf k}+{\bf q}}=\Delta_{\bf k}+\delta_{\bf q}$, where
$\delta_{\bf q}=qv_{G}\cos(\phi+\theta_{q})$ with 
$v_{G}\simeq2\Delta_{0}/k_{F}$, and 
the correction $\delta_{\bf q}$ makes a negligible contribution
to the quasiparticle energy spectrum. 
When the paramagnetic interaction is included,
$E_{{\bf k}+{\bf q}}\rightarrow E_{{\bf k}+{\bf q}}\pm \mu_{B}H$ 
in the denominator of $A({\bf k}+{\bf q},\epsilon)$,
where $\mu_{B}$ is the Bohr magneton. As $\omega\rightarrow0$ and
$T\rightarrow0$,
\begin{equation}
\sigma_{\bf q}(\omega\rightarrow0,T\rightarrow0)=
C\sum_{{\bf k},{\bf p}}t^{2}_{{\bf k}-{\bf p}}A({\bf k}+{\bf q},0)
A({\bf p},0)\;.
\end{equation}

Based on symmetry consideration,\cite{lee} one can easily deduce that
the dependence of 
$\sigma_{\bf q}(\omega\rightarrow0,T\rightarrow0)
\equiv\sigma(\theta)$ on field angle $\theta$ has a period
$\pi/2$ because of the $d$-wave gap. The sign of the gap does not matter.
Consequently, 
\begin{equation}
\sigma(\theta)=\sum_{n=0}{\cal C}_{n}
\cos(4n\theta)\;,
\end{equation}
where the ${\cal C}_{n}$'s depend on 
the ratio of the magnetic field energy
to $\gamma$ {\it i.e.},  on $E_{q}/\gamma$ and $\mu_{B}H/\gamma$.
Since both $E_{q}$ and $\mu_{B}H$ are linear to $H$,
we can parameterize the ratio $(E_{q}/\mu_{B}H)\sim d/a$. We choose
$(E_{q}/\mu_{B}H)=6$ for the
high-$T_{c}$ Bi- and Tl-based  cuprates.\cite{poole} 
Another relation can be deduced
without detailed calculation. If we expand $\sigma(\theta)$
in terms of $E_{q}/\gamma$, then its angular dependence
will appear for the first time
from the term $(E_{q}/\gamma)^{4}\cos^{4}(\phi-\theta)$
because the fourth harmonic $\cos(4\theta)$
comes from the $\cos^{4}(\theta)$ term. This means that
$\gamma < E_{q}$ is required for any clear angular dependence to show up.
For the eighth harmonic $\cos(8\theta)$ to enter,
the field has to be even higher. 

At a finite temperature $T<\gamma$, we apply the Sommerfeld 
expansion\cite{finite}
to $\sigma_{\bf q}(\omega\rightarrow0,T)\equiv\sigma(\theta,T)$
and obtain
\begin{eqnarray}
\sigma(\theta,T)=&&-C\sum_{{\bf k},{\bf p}}
t^{2}_{{\bf k}-{\bf p}}
\int{}d\epsilon\left({\partial f(\epsilon)\over\partial\epsilon}\right)
A({\bf k}+{\bf q},\epsilon)A({\bf p},\epsilon)
\nonumber\\
\simeq
&&\sigma(\theta)+{\pi^{2}\over6}
CT^{2}\sum_{{\bf k},{\bf p}}t^{2}_{{\bf k}-{\bf p}}
{\partial^{2}\over\partial\epsilon^{2}}
A({\bf k}+{\bf q},\epsilon)A({\bf p},\epsilon)\Bigr|_{\epsilon=0}
\nonumber\\
\equiv&&
\sigma(\theta)\left[1+{\pi^{2}\over6}\left({T\over\gamma}\right)^{2}
\alpha(\theta)\right]\;,
\label{finiteT}
\end{eqnarray}
where 
\begin{equation}
\alpha(\theta)=
{{\partial^{2}\over\partial\epsilon'^{2}}
\sum_{{\bf k},{\bf p}}t^{2}_{{\bf k}-{\bf p}}
A({\bf k}+{\bf q},\epsilon')A({\bf p},\epsilon')\Bigr|_{\epsilon'=0}
\over
\sum_{{\bf k},{\bf p}}t^{2}_{{\bf k}-{\bf p}}
A({\bf k}+{\bf q},0)A({\bf p},0)}\;,
\end{equation}
with $\epsilon'=\epsilon/\gamma$.

We use the nodal approximation\cite{lee,durst} to describe the low 
temperature physics 
associated with the quasiparticle transport because quasiparticles
near the nodal lines dominantly contribute to the resistive transport.
Linearizing $\xi_{\bf k}$ and $\Delta_{\bf k}$ near the nodal points on the
Fermi surface (FS), we obtain 
$\xi_{\bf k}=v_{F}({\bf k}\cdot{\hat {\bf k}}_{\perp}-k_{F})$ and
$\Delta_{\bf k}=v_{G}{\bf k}\cdot{\hat {\bf k}}_{\parallel}$, 
where $v_{F} (k_{F})$ is the Fermi
velocity (momentum) and $v_{G}=\sqrt{2}a\Delta_{0}$ is the gap velocity.
\cite{vG}
The unit vector ${\hat {\bf k}}_{\perp}$ (${\hat {\bf k}}_{\parallel}$)
is perpendicular (parallel) to the FS. These unit vectors 
will change depending on
the nodal line. For example, on the nodal line of $\phi=\pi/4$,
${\hat {\bf k}}_{\perp}=({\hat {\bf k}}_{x}+{\hat {\bf k}}_{y})/\sqrt{2}$ and
${\hat {\bf k}}_{\parallel}=(-{\hat {\bf k}}_{x}+{\hat {\bf k}}_{y})/\sqrt{2}$.
Since we include quasiparticles only near the nodal points on the FS
in our considerations,
we will use the following procedure in the actual calculation.
\begin{equation}
\sum_{k}\rightarrow
\sum_{\mbox{node}}\int{dk_{\perp}dk_{\parallel}\over{(2\pi)^{2}}}
\rightarrow
\sum_{\mbox{node}}\int{}{\cal J}dp_{1}dp_{2}\;,
\label{nodal}
\end{equation}
where $p_{1}=v_{F}k_{\perp}$, $p_{2}=v_{G}k_{\parallel}$ and
${\cal J}=\left[(2\pi)^{2}v_{F}v_{G}\right]^{-1}$. 
The coordinate transformations 
we made are rotation, translation and dilation. In the coordinate
of $(p_{1},p_{2})$, the energy dispersion of the quasiparticle $E_{\bf k}$
becomes $\sqrt{p^{2}_{1}+p^{2}_{2}} < p_{0}$, where 
$p_{0}\sim {\cal O}(\Delta_{0})$.

\section{coherent coupling}
In this section, we consider the effects of coherent coupling
on $c$-axis quasiparticle transport. 
Since $t_{{\bf k}-{\bf p}}=t_{\perp}\delta_{{\bf k}-{\bf p}}$
for a constant hopping amplitude, 
as $\omega\rightarrow0$ and $T\rightarrow0$,
$\sigma_{0}=C\sum_{\bf k}t^{2}_{\perp}A({\bf k},0)^{2}$.
For $\omega<\gamma$, we use the nodal approximation for Eq.~(\ref{finiteo}).
Then, up to order $(\omega/\gamma)^{2}$,
\begin{equation}
{\sigma(\omega)\over\sigma_{0}}=1+{1\over18}\left({\omega\over\gamma}
\right)^{2}\;,
\end{equation}
where $\sigma_{0}$ is a constant independent of $\gamma$.

In the presence of a uniform in-plane field, the quasiparticle
conductivity becomes
\begin{equation}
\sigma(\theta)=
C\sum_{\bf k}t^{2}_{\perp}A({\bf k}+{\bf q},0)
A({\bf k},0)\;.
\end{equation}
Applying the substitution Eq.~(\ref{nodal}) to the above equation, we obtain
without the paramagnetic part
\begin{eqnarray}
\sigma(\theta)=&&t^{2}_{\perp}C{\cal J}
\sum_{\mbox{node}}\int{}d\varphi\int{}
{p dp\over{p^{2}+1}}
\nonumber\\
\times
&&{1\over{p^{2}+2(E_{q}/\gamma)\cos(\phi-\theta)p\cos(\varphi)+
(E_{q}/\gamma)^{2}\cos^{2}(\phi-\theta)+1}}\;.
\end{eqnarray}
It can be shown within the nodal approximation that 
$\phi=\phi_{n}\pm\tan^{-1}\left[p_{2}/v_{G}k_{F}\right]$,
where $\phi_{n}$ is a direction on the nodal line; however,
to a good approximation, we can replace $\phi$ by $\phi_{n}$.
For the weak field case $E_{q}<\gamma$, we expand $\sigma(\theta)$
in terms of $E_{q}/\gamma$ and obtain
\begin{equation}
{\sigma(\theta)\over\sigma_{0}}\simeq
1-{1\over12}\left({E_{q}\over\gamma}\right)^{2}
+{1\over80}
\left[1-{1\over3}\cos(4\theta)\right]\left({E_{q}\over\gamma}\right)^{4}\;,
\end{equation}
where $\sigma_{0}$ is the $c$-axis quasiparticle conductivity 
in the absence of the in-plane magnetic field. 
As we see, the angular dependence
of $\sigma(\theta)$ is small for a weak field.
$\sigma(\theta)$ is maximum (minmum) when the in-plane field is along 
a nodal (anti-nodal) line. It can be physically understood as follows:
When {\bf H} is along a nodal line, for example, $\theta=\pi/4$,
the angle of ${\bf q}$ is $3\pi/4$ because ${\bf q}\propto
\left({\hat{\bf z}}\times{\bf H}\right)$. Then, while the shifted momenta
${\bf k}+{\bf q}$ of the quasiparticles with
$\phi=\pi/4$ and $\phi=5\pi/4$ deviate
from the nodal regions
when $q$ is compatible with $k$,
quasiparticles with $\phi=3\pi/4$ and $\phi=7\pi/4$
remain in the nodal areas. This means the remaining
quasiparticles govern 
the $c$-axis transport when $\theta=\pi/4$.
On the other hand, for $\theta=0$, all quasiparticles
move away from the nodal regions due to the momentum shift ${\bf q}$.
Therefore, $\sigma(\theta)<\sigma_{0}$ and $\sigma(0)<\sigma(\pi/4)$
because of a mismatch between ${\bf k}$ and ${\bf k}+{\bf q}$
in the interlayer transport.

In Fig.~1, we plot resistivity as a function of direction of the in-plane
magnetic field, $\rho(\theta)=1/\sigma(\theta)$ for various values of $E_{q}$
with (solid line) and without (dashed line)
the paramagnetic interaction included. We reproduced
results of Ref.\cite{bulaevskii} for $E_{q}/\gamma=1$, $2$, and $4$.
As shown in Fig.~1, $\rho(\theta)$ is increased as the magnetic field
increases, 
and it is decreased for a give $E_{q}$
when the paramagnetic interaction is considered
because the interaction is pair breaking. 
If we consider only the paramagnetic interaction, then
the quasiparticle conductivity $\sigma$ has no $\theta$ dependence
and increases as follows:
\begin{equation}
\sigma\rightarrow\sigma_{0}\left[1+\left({\mu_{B} H\over\gamma}\right)
\arctan\left({\mu_{B} H\over\gamma}\right)\right]\;.
\end{equation}
The eighth harmonic clearly 
appears for $E_{q}/\gamma=12$. Actually, it begins to appear 
when $E_{q}/\gamma\gtrsim10$.
Interestingly, the paramagnetic interaction
unambiguously reduces the amplitude of $\rho(\theta)$ for a higher
field $(E_{q}\gg\gamma)$. 
We also plot $\alpha(\theta)$ of Eq.~(\ref{finiteT})
for a finite $T<\gamma$ in Fig.~2. For reference in the absence of the
magnetic field,
$\alpha(\theta)=4/3$ (dashed line).

In the case of an angle-dependent hopping amplitude 
$t_{{\bf k}-{\bf p}}=t(\phi)\delta_{{\bf k}-{\bf p}}$,
the $c$-axis quasiparticle conductivity in zero field is
$\sigma_{0}=C\sum_{\bf k}t(\phi)^{2}A({\bf k},0)^{2}$.
For $\omega<\gamma$, using the nodal approximation we obtain
\begin{equation}
{\sigma(\omega)\over\sigma_{0}}=1+{34\over9}\ln\left({\Delta_{0}\over
\gamma}\right)\left({\omega\over\Delta_{0}}\right)^{2}\;,
\end{equation}
where $\sigma_{0}\propto(\gamma/\Delta_{0})^{2}$ in this case.
Effects of the in-plane field on the $c$-axis quasiparticle conductivity
can be seen in the same way as before, namely,
\begin{equation}
\sigma(\theta)=
C\sum_{\bf k}t(\phi)^{2}A({\bf k}+{\bf q},0)
A({\bf k},0)\;.
\end{equation}
Since $t(\phi)=t_{\phi}(p_{2}/\Delta_{0})^{2}$ within the nodal
approximation, $\sigma(\theta)$ becomes without the paramagnetic 
interaction
\begin{eqnarray}
\sigma(\theta)=&&t^{2}_{\phi}C{\cal J}
\sum_{\mbox{node}}\int{}d\varphi\int{}
{p dp\over{p^{2}+1}}\left({p\over\Delta_{0}}\right)^{4}\sin^{4}(\varphi)
\nonumber\\
\times
&&{1\over{p^{2}+2(E_{q}/\gamma)\cos(\phi-\theta)p\cos(\varphi)+
(E_{q}/\gamma)^{2}\cos^{2}(\phi-\theta)+1}}\;.
\end{eqnarray}
In this case, we find that $\sigma(\theta)/\sigma_{0}\simeq1$ 
without clear dependence on the direction of the in-plane field, and
$\alpha(\theta)\sim {\cal O}\left[(\gamma/\Delta_{0})^{2}
\ln(\Delta_{0}/\gamma)\right]$ in the leading order and the next order
is ${\cal O}\left[(E_{q}/\Delta_{0})^{2}\right]$; therefore,
$\alpha(\theta)$ is negligible.
The $c$-axis quasiparticle conductivity
is insensitive to the in-plane magnetic field
because the angle-dependent hopping amplitude $t_{\perp}(\phi)$
eliminates contributions of nodal quasiparticles, which otherwise manifest
the effects of the field.

\section{incoherent coupling}
For impurity-mediated incoherent coupling $t_{{\bf k}-{\bf p}}=
V_{{\bf k}-{\bf p}}$, 
we need a model
for the impurity scattering potential and need to carry out an average over
impurity configurations. We use a simple model\cite{hirschfeld} for
the scattering potential $|V_{{\bf k}-{\bf p}}|^{2}=
|V_{0}|^{2}+|V_{1}|^{2}\cos(2\phi_{k})\cos(2\phi_{p})$.
One may expand $|V_{{\bf k}-{\bf p}}|^{2}$ with respect to
scattering symmetry so that it is decomposed
\begin{equation}
|V_{{\bf k}-{\bf p}}|^{2}=|V_{0}|^{2}
+\sum_{l,l'}|V_{ll'}|^{2}\cos[(4l+2)\phi_{k}]\cos[(4l'+2)\phi_{p}]\;,
\end{equation}
where $l$ and $l'$ are integers. However, for the $c$-axis
quasiparticle transport, only the $|V_{0}|^{2}$ term
gives a non-zero value to the conductivity.
As $\omega\rightarrow0$ and $T\rightarrow0$,
$\sigma_{0}=C|V_{0}|^2\sum_{{\bf k},{\bf p}}A({\bf k},0)
A({\bf p},0)$. For $\omega<\gamma$, it can be shown that
\begin{equation}
{\sigma(\omega)\over\sigma_{0}}=1+{1\over3}{(\omega/\gamma)^{2}\over
\ln(\Delta_{0}/\gamma)}\;,
\end{equation}
where $\sigma_{0}\propto
[\gamma\ln(\Delta_{0}/\gamma)/\Delta_{0}]^{2}$ for impurity-mediated
incoherent coupling. 

In the presence of an in-plane magnetic field,
the $c$-axis quasiparticle conductivity becomes
\begin{equation}
\sigma(\theta)=
C|V_{0}|^2\sum_{{\bf k},{\bf p}}A({\bf k}+{\bf q},0)
A({\bf p},0)\;.
\end{equation}
As we see, the angular dependence of $\sigma(\theta)$ is determined
by $\sum_{\bf k}A({\bf k}+{\bf q},0)$. It can be shown numerically
and analytically that
the field effect appears only as the paramagnetic contribution.
We present some analytic results, for example, 
in the weak field limit
\begin{equation}
\sum_{\bf k}\Bigl[A({\bf k}+{\bf q},0)-A({\bf k},0)\Bigr]\simeq
{2\gamma\over{\pi v_{G}v_{F}}}\;h\left({\mu_{B}H\over\gamma}\right)\;,
\end{equation}
where $h(x)=x^{2}-x^{4}/6+x^{6}/15$. Therefore, the $c$-axis
quasiparticle conductivity for a weak field becomes
\begin{equation}
{\sigma(\theta)\over\sigma_{0}}\simeq
1+{h(\mu_{B}H/\gamma)\over\ln(\Delta_{0}/\gamma)}\;.
\end{equation}
Consequently, for incoherent coupling the in-plane magnetic field
has no effect on the $c$-axis quasiparticle transport if
the paramagnetic interaction is not considered. 
Physically, the momentum of the transport quasiparticle is not
constraint in incoherent coupling
so that the change in momentum ${\bf q}$ does not matter.
Since the paramagnetic 
interaction is pair breaking, the magnitude of $\sigma(\theta)$
is increased but shows no angular dependence when this interaction
is included.
At a finite temperature $(T<\gamma)$, we use Eq.~(\ref{finiteT})
and for a weak field, $\alpha(\theta)$ becomes
\begin{equation}
\alpha(\theta)\simeq
1+{(\pi^{2}/3)\over{\left[\ln(\Delta_{0}/\gamma)+h(\mu_{B}H/\gamma)\right]
\left[1-(\mu_{B}H/\gamma)^{2}\right]}}\;.
\end{equation}

\section{$c$-axis conductivity sum rule}
From the $c$-axis conductivity sum rule, the superfluid density $\rho_{s}$
can be written in terms of the missing spectral weight $(N_{n}-N_{s})$
and
the theral averages of kinetic energy $\langle H_{c}\rangle^{s(n)}$
of a superconducting (normal) state as follows:
\begin{equation}
\rho_{s}=(N_{n}-N_{s})-4\pi e^{2}d
\Bigl[\langle H_{c}\rangle^{s}-\langle H_{c}\rangle^{n}\Bigr],
\end{equation}
where
$\omega_{c}$ is the cutoff frequency for
the interband transitions that $H_{c}$
does not describe. Using the above equation as well as
the Kramers-Kronig relation between the conductivity and the penetration
depth, we obtain the normalized missing spectral weight
$(N_{n}-N_{s})/\rho_{s}$;
\begin{equation}
{(N_{n}-N_{s})\over\rho_{s}}={1\over2}+
{1\over2}{\sum_{\omega}\sum_{{\bf k},{\bf p}}
|t_{{\bf k}-{\bf p}}|^{2}[G({\bf k},\omega)G({\bf p},\omega)-
G_{0}({\bf k},\omega)G_{0}({\bf p},\omega)]\over
\sum_{\omega}\sum_{{\bf k},{\bf p}}
|t_{{\bf k}-{\bf p}}|^{2}F({\bf k},\omega)F^{+}({\bf p},\omega)}\;,
\end{equation}
where $G({\bf k},\omega)$ and $F({\bf k},\omega)$ are superconducting
Green functions and $G_{0}({\bf k},\omega)$ is in the normal state.
\cite{kim1,kim2}

Since, in principle, all possible interlayer
couplings might be present in a given sample,
we need to consider coherent and incoherent
coupling at the same time.
The actual $c$-axis transport of quasiparticles is determined from
competing effects between couplings. Based on the conductivity sum rule,
we can estimate the ratio
$(\sigma_{co}/\sigma_{in})$, where $\sigma_{co}$ $(\sigma_{in})$
is the $c$-axis
quasiparticle conductivity for coherent (incoherent) coupling.
Since the sum rule does not distinguish between $t_{\perp}$ and $t(\phi)$,
we consider two limiting cases: i) $t_{{\bf k}-{\bf p}}=
t_{\perp}\delta_{{\bf k}-{\bf p}}+V_{{\bf k}-{\bf p}}$ and ii)
$t_{{\bf k}-{\bf p}}=t(\phi)\delta_{{\bf k}-{\bf p}}+V_{{\bf k}-{\bf p}}$.
As $\omega\rightarrow0$ and $T\rightarrow0$, for the case with $t_{\perp}$
\begin{eqnarray}
\sigma=&&\sigma_{co}+\sigma_{in}
=\sigma_{co}\left[1+{\sigma_{in}\over\sigma_{co}}\right]
\nonumber\\
=&&\sigma_{co}\left[1+{4\Delta_{0}\sigma_{cn}\over
\pi e^{2}dt^{2}_{\perp}N(0)}\left({\gamma\over\Delta_{0}}
\ln(\Delta_{0}/\gamma)\right)^{2}\right]\;,
\end{eqnarray}
where we have used $\sigma_{co}=2e^{2}dt^{2}_{\perp}N(0)/(\pi\Delta_{0})$
and $\sigma_{in}=(8/\pi^{2})\sigma_{cn}\left[\gamma\ln(\Delta_{0}/
\gamma)/\Delta_{0}\right]^{2}$ with $\sigma_{cn}=4\pi n_{i}d(eV_{0}N(0))^{2}$.
One may think that we need to know the ratio
of $|V_{0}|^{2}$ to $t^{2}_{\perp}$ in order to compare $\sigma_{co}$
with $\sigma_{in}$.
However, the conductivity sum rule helps us to estimate the ratio
$(\sigma_{co}/\sigma_{in})$ without {\it {ad hoc}} information.

Let us define $\eta=\rho_{s,co}/\rho_{s,in}=(1/\lambda^{2}_{co})/
(1/\lambda^{2}_{in})$, where $\rho_{s}$ and $\lambda$ are the
corresponding superfluid density and penetration depth, respectively.
For coherent coupling
\begin{equation}
{1\over\lambda^{2}_{co}}= 16\pi e^{2}dt^{2}_{\perp}N(0)
\left[{1\over2}-{1\over\pi}{\bf K}(i\Delta_{0}/\gamma)\right]\;,
\end{equation}
where ${\bf K}$ is the complete elliptic integral of the first kind.
For incoherent coupling, we assume $|V_{0}|^{2}=|V_{1}|^{2}$ for simplicity
and to illustrate possible effects.
Then, 
\begin{equation}
{1\over\lambda^{2}_{in}}= 32\sigma_{cn}
\sum_{\omega}\left[(\kappa'^{2}/\kappa){\bf K}(\kappa)
-{\bf E}(\kappa)/\kappa\right]^{2}\;,
\end{equation}
where $\kappa=\Delta_{0}/\sqrt{\Delta^{2}_{0}
+(\omega+\gamma\mbox{sgn}\omega)^{2}}$,
$\kappa'=\sqrt{1-\kappa^{2}}$,
and ${\bf E}$ is
the complete elliptic integral of the second kind.\cite{kim2}
Since $\gamma << \Delta_{0}$, 
$\eta\simeq 8\pi e^{2}dt^{2}_{\perp}N(0)/
[12\Delta_{0}\sigma_{cn}]$ and we obtain
\begin{equation}
\sigma=\sigma_{co}\left[1+{(8/3)\over\eta}
\left({\gamma\over\Delta_{0}}\ln(\Delta_{0}/\gamma)\right)^{2}
\right]\;.
\label{cond1}
\end{equation}
Since the total superfluid density $\rho_{s}=\rho_{s,co}+\rho_{s,in}$,
it can be shown that
\begin{equation}
{N_{n}-N_{s}\over\rho_{s}}\simeq
{1+2\eta\over2(1+\eta)}+{1.08\over 1+\eta}\;.
\label{sum}
\end{equation}
Note that $\eta$ in Eq.~(\ref{cond1}) can be determined from the violation of
the conductivity sum rule Eq.~(\ref{sum}).
$\eta\rightarrow\infty$ corresponds to pure coherent $c$-axis coupling
in which case the sum rule Eq.~(\ref{sum}) is conventional and equal to $1$.
On the other hand, $\eta\rightarrow0$ means pure incoherent coupling
and the sum rule is larger than $1$ and equal to $1.58$ in our model
with $|V_{0}/V_{1}|=1$. A finite $\eta$ corresponds to an admixture
of coherent and incoherent $c$-axis transport. For example,
$\eta\simeq5$ applies when $(N_{n}-N_{s})/\rho_{s}=1.1$, that is,
there is a $10\%$ violation of the sum rule upwards.
The sum rule itself determines the admixture of $\rho_{s,co}$ to
$\rho_{s,in}$, but cannot differentiate between $t_{\perp}$ and
$t(\phi)$ for the coherent part. Limits on the relative size 
of these two overlap integrals can only be set from consideration
of the chemistry of the CuO$_{2}$ planes and their overlap or, 
alternatively, from experimental information such as the behavior of
$\sigma(0,0)$ when the impurity content is increased.

For the case with $t(\phi)$
\begin{equation}
{1\over\lambda^{2}_{co}}\simeq 16\pi e^{2}dt^{2}_{\phi}N(0)
\left[{3\over16}-{2\over3\pi}
{\gamma\over\sqrt{\gamma^{2}+\Delta^{2}_{0}}}
{\bf E}\left({\Delta_{0}\over\sqrt{\gamma^{2}+\Delta^{2}_{0}}}
\right)\right]\;,
\end{equation}
where ${\bf E}$ is the complete elliptic integral of the second kind,
and $\eta$ as above with $3\leftrightarrow8$ and 
$t_{\phi}\leftrightarrow t_{\perp}$.
Following a similar way as before, we obtain
\begin{equation}
\sigma=\sigma^{\prime}_{co}\left[1+{(8/3)\over\eta}
\ln^{2}(\Delta_{0}/\gamma)
\right]\;,
\label{cond2}
\end{equation}
where $\sigma^{\prime}_{co}=\sigma_{co}$
with $(3/4)\leftrightarrow2$ and
$t_{\phi}\leftrightarrow t_{\perp}$.

Note that $t_{\perp}$ and $t_{\phi}$ case both contribute in the same order
to the zero temperature $c$-axis superfluid density. For the pure case,
assuming $t_{\perp}$ and $t_{\phi}$ are of the same magnitude, the ratio of 
the two contributions is $8/3$. This is in sharp contrast to what
we found in the case of $\sigma(0,0)$ where $t_{\perp}$ contributes
of order $t^{2}_{\perp}$ while the contribution from $t_{\phi}$
is of order $t^{2}_{\phi}(\gamma/\Delta_{0})^{2}$ and so vanishes
in the clean limit.
Recently, Gaifullin {\it et al.}\cite{gaif} have measured the temperature
dependence of the $c$-axis superfluid density in Bi$2212$ 
and found that at low
temperature it is well fit by a form $\left[1-A(T/T_{c})^{\alpha}\right]$
with $\alpha$ of order $4-6$ close to the values reported for
Hg$1201$.\cite{panago} This finding favours a pure $t(\phi)$ model
which is known to give a $T^{5}$ law (Ref.\cite{xiang}) 
and leaves little room for a
subdominant $t_{\perp}$ contribution because in this case the $c$-axis
penetration depth will mirror perfectly\cite{radtke}
its in-plane temperature
dependence which goes like $T$.
The data certainly preclude a linear $T$ contribution to the superfluid
density of more than a few percent implying a ratio $(t_{\perp}/t_{\phi})$
of order $< 10^{-1}$ at most. In this instance, the ratio
$(\gamma/\Delta_{0})$ can easily be comparable to $(t_{\perp}/t_{\phi})$
and can even be the larger of the two. 
As an illustration in the experimental work
establishing the in-plane universal limit for the thermal
conductivity\cite{univ} the values of $(\gamma/\Delta_{0})$ are of
order $\lesssim 10^{-1}$. Latyshev {\it et al.}\cite{latyshev}
quoted $\gamma\approx 3$meV for their single crystals, which is of similar 
order. This implies that in the experiments on Bi$2212$,
the contribution to $\sigma(0,0)$ of each of $t_{\perp}$ and $t_{\phi}$
terms are likely to be close in magnitude.
Nevertheless, it is concluded in Ref.\cite{latyshev} that the universal
limit is observed so that $t_{\perp}$ presumably still dominates.
Additional work with various values of $\gamma$, 
somewhat larger as well as smaller than used so far, should allow 
one to establish the size of the important ratio $(t_{\perp}/t_{\phi})$.
This information, with the $c$-axis optical sum rule, would allow
an estimate of each of the three possible contributions to the $c$-axis
transport discussed in this paper. These parameters are important because
they control all $c$-axis transport.
We point out here that we have considered separately the case
$t_{\perp}$ and $t(\phi)$. If amplitudes are added and squared, a cross
term of the form $2t_{\perp}t_{\phi}\cos^{2}(2\phi)$ will enter but
this will not change qualitatively the conclusion made above.
The contribution of such a term to $\sigma_{0}$ is 
$(4e^{2}dN(0)/\pi\Delta_{0})
t_{\perp}t_{\phi}(\gamma/\Delta_{0})^{2}\ln(\Delta_{0}/\gamma)$
and to the superfluid density is
$16\pi^{2}e^{2}dN(0)t_{\perp}t_{\phi}
\left[1/2-(2/\pi)\Bigl(\gamma/\sqrt{\gamma^{2}+\Delta^{2}_{0}}\Bigr)
{\bf E}\Bigl(\gamma/\sqrt{\gamma^{2}+\Delta^{2}_{0}}\Bigr)\right]$.

\section{conclusions}
We have discussed how the nature of the interlayer coupling influences
the $c$-axis quasiparticle conductivity $\sigma_{\bf q}(\omega,T)$
in the absence and in the presence of an in-plane magnetic field.
In zero field, $\sigma(0,0)$ is independent of the in-plane scattering rate
$\gamma$, to leading order, only for coherent coupling with a constant
hopping amplitude. Also,
$\sigma(\omega,0)$ shows a different behavior for different $c$-axis
coupling.
Similarly, the field effects depend crucially
on the nature of the interalyer coupling. For coherent coupling with
a constant hopping amplitude, an angular dependence of the conductivity
$\sigma(\theta)$
appears in high
field. The resistivity $\rho(\theta)$ increases because the mismatch
between ${\bf k}$ and ${\bf k}+{\bf q}$ decreases $\sigma(\theta)$, and
$\rho(\theta)$ is maximum (minimum) when the field is along
the anti-nodal (nodal) line. This confirms previous work.\cite{bulaevskii}
When the paramagnetic interaction is included,
$\rho(\theta)$ is decreased because this interaction is pair breaking
and the anisotropy also decreases.
For high field $(E_{q}\gg\gamma)$,
the eighth harmonic appears and can be seen in the $\theta$ dependence
of $\rho(\theta)$.
For coherent coupling with an angle-dependent hopping
amplitude, $\sigma(\theta)$ is insensitive to field direction 
because this hopping amplitute eliminates the contribution
of the nodal quasiparticles which manifest the angular dependence.
Also, the paramagnetic interaction does not significantly
change $\sigma(\theta)$.
For incoherent coupling, only the paramagnetic interaction
has an effect on the $c$-axis quasiparticle transport and
$\sigma(\theta)$ has no dependence on field direction.
The coefficient of the first temperature correction which goes like $T^{2}$
is anisotropic only in the constant $t_{\perp}$ case and shows 
the same minima (maxima)
as does the leading contribution.

The $c$-axis conductivity sum rule helps estimate separately the
contributions from coherent and incoherent coupling to the
quasiparticle transport without microscopic information
on hopping amplitudes. 
However, it cannot, by itself, differentiate between contributions
from constant or angular dependent coherent hopping amplitude.
To get separate information on these two contributions, impurity
studies could be used. Consideration of the temperature dependence of the 
$c$-axis superfluid density measured in Josephson plasma resonance 
experiments already gives evidence that $t_{\phi}$ is much larger than
$t_{\perp}$, but do not unambiguously rule out a small $t_{\perp}$
contribution which could still be large enough to dominate the universal
limit for $\sigma(0,0,)$ in relatively pure samples 
$(t_{\perp}/t_{\phi}\;>\;\gamma/\Delta_{0})$.
At the same time, $t_{\phi}$ would dominate the temperature dependence
of the superfluid density.
We mention that the sum rule of Eq.~(\ref{sum})
was derived under the assumption
that the normal state is Fermi-liquid like. However,
if it has a non-Fermi-liquid nature such as a pseudogap, the sum rule
has to be modified to account for this; 
therefore, it is necessary
to explore the competing effects of interlayer coupling in a more fundamental
theory which goes beyond a Fermi liquid normal state.

Research was supported in part by the Natural Science and Engineering
Research Council of Canada (NSERC) and by the Canadian Institute
for Advanced Research (CIAR).
W. K. thanks E. H. Kim for many useful discussions.

\begin{figure}

\caption{For various values of $E_{q}/\gamma(=1,4,6,12)$, resistivity 
$\rho(\theta)/\rho_{0}=\sigma_{0}/\sigma(\theta)$ 
as a function of a direction of 
the in-plane field $\theta$ is plotted 
with (solid curve) and without (dashed curve) the paramagnetic interaction. 
As $E_{q}/\gamma$ is increased, $\rho(\theta)/\rho_{0}$ is increased, 
and for
a high field $(E_{q}/\gamma=12)$,
the paramagnetic interaction unambiguously reduces the amplitude of
$\rho(\theta)$ and its anisotropy}
\vskip 0.5in
\caption{$\alpha(\theta)$ in Eq.~(\ref{finiteT}) shows
finite temperature effects on the $c$-axis qusiparticle conductivity
for $E_{q}/\gamma=1$, $2$, and $3$. The paramagnetic interaction is
included.
In the absence of the field, $\alpha(\theta)=4/3$ (dahsed line).}

\end{figure}


\begin{references}
\bibitem{vG} It can be shown that $v_{G}\propto\Delta_{0}/k_{F}$
in the continuum limit.
\end{references}
\end{document}